\documentclass[aps,floats,twocolumn,showpacs,showkeys]{revtex4}
\usepackage{graphicx}
\usepackage{epsfig}
\input{psfig.sty}

\newcommand{\bastar}{\begin{eqnarray*}}
\newcommand{\eastar}{\end{eqnarray*}}
\newskip\humongous \humongous=0pt plus 1000pt minus 1000pt

\newif\ifdtup

\relax
\newcommand{\be}{\begin{equation}}
\newcommand{\ee}{\end{equation}}
\newcommand{\bea}{\begin{eqnarray}}
\newcommand{\eea}{\end{eqnarray}}
\newcommand{\pro}{\partial}
\newcommand{\n}{\hat n}
\newcommand{\oneg}{\displaystyle\frac{1}{g}}
\newcommand{\D}{{\hat D}}

\newcommand{\X}{{\vec X}}
\newcommand{\vX}{{\vec X}}

\newcommand{\hn}{\hat n}

\newcommand{\dfrac}{\displaystyle\frac}
\newcommand{\ba}{\begin{array}}
\newcommand{\ea}{\end{array}}
\newcommand{\nn}{\nonumber}

\newcommand{\valpha}{{\vec \alpha}}

\begin{document}
\title{Effective Potential of QCD}
\bigskip
\author{Y.M. Cho and J.H. Kim}
\email{ymcho@yongmin.snu.ac.kr}
\affiliation{Center for Theoretical Physics
and School of Physics, \\
College of Natural Sciences,
Seoul  National University, Seoul 151-742, Korea}
\author{D.G. Pak}
\email{dmipak@phya.snu.ac.kr}
\affiliation{ Center for Theoretical Physics, 
Seoul National University, Seoul 151-742, Korea\\
Institute of Applied Physics, 
Uzbekistan National University, Tashkent 700-095, Uzbekistan}
                                                                                
\begin{abstract}
We present a simple method to calculate the one-loop effective 
action of QCD which reduces the calculation to that of $SU(2)$ QCD.
For the chromomagnetic background we show that the effective 
potential has an absolute minimum only when two color magnetic fields 
$H_{\mu\nu}^3$ and $H_{\mu\nu}^8$ are orthogonal to each other. 
For the chromoelectric background we find that 
the imaginary part of the effective action has a negative signature,
which implies the gluon pair-annihilation. We discuss 
the physical implications of our result.

\end{abstract}
                                                                                
\maketitle
                                                                                
\section{Introduction}
                                                                                
An interesting problem which has been studied recently 
is to calculate the soft gluon production
rate in a constant chromoelectric background \cite{nayak,plb05}.
This problem is closely related to the problem to calculate 
the QCD effective action in a constant chromoelectric background,
because the imaginary part of the effective action
determines the production rate \cite{schw,prl01}.
This leads to one of the most outstanding problems
in theoretical physics, the problem to calculate the QCD 
effective action in an arbitrary homogeneous (constant) chromomagnetic 
and chromoelectric background, which has been studied 
by many authors \cite{savv,niel,yil,adl,sch,prd02,jhep}.
The purpose of this Letter is to repeat the calculation 
with a different method, 
and to study the vacuum structure of the effective action. 

For the chromomagnetic background we find that 
the effective potential has a non-trivial dependence
on the relative space orientation of two magnetic fields 
$H_{\mu\nu}^3$ and $H_{\mu\nu}^8$.
Given the fact that the classical potential depends only on
$(H_{\mu\nu}^3)^2+(H_{\mu\nu}^8)^2$, this is surprising.
More importantly, it has an absolute minimum only when 
they become orthogonal to each other. 
When they are parallel, the effective potential has two degenerate 
local minima. To the best of our knowledge, this constitutes 
a new evidence that a quantum fluctuation can actually 
determine the space orientation of a magnetic field.
For the chromoelectric background we find that the imaginary 
part of the effective action become negative, 
which should be contrasted with known results \cite{yil,sch}. 

To calculate the one-loop effective action one must decompose
the gluon field into two parts, the slow-varying
classical background $\vec B_\mu$ and the fluctuating quantum
part $\vec Q_\mu$, and integrate the quantum part \cite{dewitt,pesk}.
But this decomposition has to be gauge independent 
for the effective action to be
gauge independent. A natural way to have a gauge independent
decomposition is to make the Abelian projection. 
To make the Abelian projection we let $\n$ be the color 
octet unit vector which selects the color direction at each 
space-time point, and require \cite{cho1,cho2}
\bea
D_\mu \n =0.~~~~~(\n^2 =1)
\label{ap}
\eea
This generates another constraint on the gauge potential
\bea
&D_\mu \n' =0,~~~\n'^c = \sqrt 3 d_{ab}^{~~c} \n^a \n^b. 
~~~(\n'^2=1)
\label{n'}
\eea
Notice that when $\n$ is $\lambda_3$-like, $\n'$ becomes
$\lambda_8$-like, so that one may always choose $\n$ to be
a $\lambda_3$-like $\n_3$ and $\n'$ a $\lambda_8$-like $\n_8$. 
The Abelian projection uniquely determine the restricted potential,
the most general Abelian gauge potential $\hat A_\mu$, in QCD
\bea
&\hat A_\mu= \sum_p \Big(A_\mu^i \n_i 
- \oneg \n_i\times\pro_\mu\n_i \Big).~~~(i=3,8) 
\label{rp}
\eea
where $A_\mu^i=\n_i\cdot \vec A_\mu$ are the chromoelectric potentials.
With this the most general QCD potential is written as
\bea
&\vec{A}_\mu = \hat A_\mu + \X_\mu, 
~~~\hat{n_i}\cdot\vec{X}_\mu=0,~~~(i=3,8)
\label{cdec}
\eea
where $\X_\mu$ is the valence potential. 

The decomposition (\ref{cdec}) allows two types of gauge 
transformation, the background gauge transformation described by
\bea
&\delta \hat A_\mu = \oneg \D_\mu \valpha,
~~~~~\delta \X_\mu = - \valpha \times \X_\mu, 
\label{bgt}
\eea
and the quantum gauge transformation described by
\bea
&\delta \hat A_\mu = 0,
~~~~~&\delta \X_\mu =\oneg \D_\mu \valpha.
\label{qgt}
\eea
The background gauge transformation shows that $\hat A_\mu$ 
by itself enjoys the full $SU(3)$ 
gauge degrees of freedom, even though it describes the Abelian
part of the potential. Furthermore
$\vec X_\mu$ transforms covariantly under (\ref{bgt}),
which is why we call it the valence potential.
But what is most important is that the decomposition (\ref{cdec})
is gauge-independent. 
Once the color direction $\hn$ is selected, the
decomposition follows automatically,
independent of the choice of a gauge \cite{cho1,cho2}.
                                                                                
With the decomposition (\ref{cdec}) one has
\bea
&\vec{F}_{\mu\nu}=\hat F_{\mu \nu} + \D _\mu \vX_\nu -
\D_\nu \vX_\mu + g\vX_\mu \times \vX_\nu, \nn\\
&\hat F_{\mu \nu}= G_{\mu\nu}^i \hn_i,
~~~~~G_{\mu\nu}^i =F_{\mu \nu}^i+ H_{\mu \nu}^i \nn\\
&F_{\mu \nu}^i= \pro_\mu A_\nu^i -\pro_\nu A_\mu^i, \nn\\
&H_{\mu \nu}^i=-\dfrac{1}{g} \hn_i \cdot (\pro_\mu \hn_i \times
\pro_\nu \hn_i)= \pro_\mu C_\nu^i -\pro_\nu C_\mu^i,
\label{fmn}
\eea
where $C_\mu^i$ is the chromomagnetic potentials.
This tells that restricted potential has a dual structure.
With (\ref{fmn}) the QCD Lagrangian can be written as follows
\bea
&{\cal L} =-\dfrac{1}{4} {\hat F}_{\mu\nu}^2 
-\dfrac{1}{4}(\D_\mu\vX_\nu-\D_\nu\vX_\mu)^2 \nn \\
&-\dfrac{g}{2} {\hat F}_{\mu\nu} \cdot (\vX_\mu \times \vX_\nu)
-\dfrac{g^2}{4} (\vX_\mu \times \vX_\nu)^2.
\eea
This shows that QCD is a restricted gauge theory which has
a gauge covariant valence gluon as a colored source.
                                                                                
With this we can integrate  
the quantum field $\vec X_\mu$ with the gauge fixing
$\hat D_\mu \vec X_\mu = 0$. 
For this we first introduce
three complex vector fields ($W_\mu^p,~p=1,2,3$ )
\begin{widetext}
\bea
& W^{1}_{\mu}= \dfrac{1}{\sqrt 2} (X_\mu^1 + i X_\mu^2),
~~~W^{2}_{\mu}= \dfrac{1}{\sqrt 2} (X_\mu^6 + i X_\mu^7),
~~~W^{3}_{\mu}= \dfrac{1}{\sqrt 2} (X_\mu^4 - i X_\mu^5),
\eea
and express the QCD Lagrangian as
\bea 
&{\cal L} = -\dfrac{1}{6} \sum_p ({\cal G}_{\mu\nu}^p)^2 
+\dfrac{1}{2} \sum_p |D_{p\mu} W_{\nu}^p- D_{p\nu} W_{\mu}^p|^2 
+ig \sum_p {\cal G}_{\mu\nu}^p W_\mu^{p*} W_\nu^p 
-\dfrac{1}{2} g^2 \sum_p \big[(W_\mu^{p*}W_\mu^p)^2
-(W_\mu^{p*})^2 (W_\nu^p)^2 \big] , \nn\\ 
&{\cal G}_{\mu\nu}^p = \partial_\mu {\cal B}_\nu^p-
\partial_\nu {\cal B}_\mu^p, 
~~~~~D_{p\mu}W_\nu^p = (\partial_\mu + ig {\cal B}_\mu^p) W_\nu^p,  \nn\\
&{\cal B}_\mu^1= B_\mu^3,~~~{\cal B}_\mu^2=-\dfrac{1}{2}B_\mu^3
+\dfrac{\sqrt 3}{2}B_\mu^8,
~~~{\cal B}_\mu^3=-\dfrac{1}{2}B_\mu^3
-\dfrac{\sqrt 3}{2}B_\mu^8,
~~~B_\mu^i=A_\mu^i+C_\mu^i,
\eea
Notice that the potentials ${\cal B}_\mu^p$ are precisely the 
dual potentials in $i$-spin, $u$-spin, and $v$-spin 
direction in color space which couple to three valence gluons $W_{\mu}^p$.
With this we have the following integral expression of 
the one-loop effective action
\bea
&\exp \Big[iS_{eff}(\hat A_\mu) \Big] = \dfrac{}{} \sum_p \int
{\cal D} W_\mu {\cal D} W_\mu^* {\cal D}c_1{\cal D}c_1^{\dagger}
{\cal D}c_2{\cal D}c_2^{\dagger} \exp \Big{\lbrace} \dfrac{}{}
i\int \Big[ -\dfrac{1}{6} {\cal G}_{\mu\nu}^2
-\dfrac{1}{2}|{D}_\mu{W}_\nu-{D}_\nu{W}_\mu|^2 \nn\\
&+ig {\cal G}_{\mu\nu} W_\mu^* W_\nu 
-\dfrac{1}{2} g^2 \Big[(W_\mu^*W_\mu)^2-(W_\mu^*)^2 (W_\nu)^2 \Big] 
-\dfrac{1}{\xi} |{D}_\mu W_\mu|^2 \nn\\
&+ c_1^\dagger ({D}^2 + g^2W_\mu^* W_\mu )c_1 - g^2 c_1^\dagger
 W_\mu W_\mu c_2 + c_2^{\dagger}
({D}^2 + g^2W_\mu^* W_\mu )^*c_2
- g^2 c_2^{\dagger} W^*_\mu W^*_\mu c_1 ~\Big] d^4x \Big{\rbrace},
\label{ea1}
\eea
where $\vec c$ and ${\vec c}^{~*}$ are the ghost fields.
Notice that here we have suppressed the summation index $p$ 
in the integrand. Now a few remarks are in order. First,
notice that except for the $p$-summation the integral expression is 
identical to that of $SU(2)$ QCD \cite{prd02,jhep}.
This shows that one can reduce the calculation of QCD
effective action to that of $SU(2)$ QCD.
Secondly, the above result can easily be 
generalized to $SU(N)$ QCD with $N(N-1)/2$ $p$-summation.
Thirdy, one might include the Abelian part in the functional 
integration, but this does not affect the result because 
the Abelian part has no self-interaction. 
This tells that only the valence gluon loops contribute
to the integration. Finally, the bakground $\hat F_{\mu\nu}$
can still have a non-trivial fluctuation, because
$\hn_3$ and $\hn_8$ have an arbitrary space-time dependence.
Only ${\cal G}_{\mu\nu}$ need be constant \cite{cho3}.

Now, from the $SU(2)$ QCD result we have \cite{prd02,jhep}
\bea
&\Delta S = i \sum_p \ln {\rm Det} (-D_p^2+2gH_p)
(-D_p^2-2gH_p)
+ i \sum_p \ln {\rm Det} (-D_p^2-2igE_p)(-D_p^2+2igE_p)
- 2i \sum_p \ln {\rm Det}(-D_p^2), \nn\\
&H_p = \dfrac{1}{2} \sqrt {\sqrt {{\cal G}_p^4
+ ({\cal G}_p \tilde {\cal G}_p)^2} + {\cal G}_p^2},
~~~~~E_p = \dfrac{1}{2} \sqrt {\sqrt {{\cal G}_p^4
+ ({\cal G}_p \tilde {\cal G}_p)^2} - {\cal G}_p^2},
\label{fdabx}
\eea
from which we obtain
\bea
&\Delta {\cal L} =  \lim_{\epsilon\rightarrow0}
\dfrac{g^2}{16 \pi^2} \sum_p  \int_{0}^{\infty}
\dfrac{dt}{t^{3-\epsilon}} \dfrac{H_p E_p t^2}{\sinh (gH_p t/\mu^2)
\sin (gE_p t/\mu^2)} \Big[ \exp(-2gH_p t/\mu^2)+\exp(+2gH_p t/\mu^2) \nn\\
&+\exp(+2igE_p t/\mu^2)+\exp(-2igE_p t/\mu^2)-2 \Big].
\label{eaabx}
\eea
\end{widetext}
Notice that for the magnetic background we have $E_p=0$,
but for the electric background we have $H_p=0$.

The evaluation of the funtional integral is straightforward.
But it has the well-known infra-red divergence which has 
to be regularized, and the funtional integration depends on
the regularization method \cite{savv,niel,yil,adl,sch,prd02,jhep}. 
Consider the magnetic background first. 
With the $\zeta$-function regularization 
we obtain \cite{niel,prd02,jhep,cho3}
\bea
&{\cal L}_{eff} = - \sum_p \Big(\dfrac{H_p^2}{3} 
+\dfrac{11g^2}{48\pi^2} H_p^2(\ln \dfrac{gH_p}{\mu^2}-c) \nn\\
&+ i \dfrac{g^2}{8\pi} H_p^2 \Big).
\label{seaa}
\eea
But if we adopt the gauge invariant regularization which 
respects the causality,
we find that the effective action has the same real part,
but no imaginary part \cite{sch,prd02,jhep,cho3}

As for the electric background we find, using 
the $\zeta$-function regularization \cite{prd02,jhep,cho3}, 
\bea
&{\cal L}_{eff} = \sum_p \Big(\dfrac{E_p^2}{3}
+\dfrac{11g^2}{48\pi^2} E_p^2(\ln \dfrac{gE_p}{\mu^2}-c) \nn\\
&- i \dfrac{23g^2}{96\pi} E_p^2 \Big).
\label{seab}
\eea
But with the gauge invariant regularization we find
that the imaginary part changes to \cite{sch,prd02,jhep,cho3}
\bea
\label{ceab}
& Im~{\cal L}_{eff} = - \sum_p \dfrac{11g^2}{96\pi} E_p^2.
\eea
Notice that, independent of the regularization method, 
the imaginary part has a negative
signature. This might look strange, because
this implies a negative probability of gluon pair-creation.
But we notice that the negative signature is a direct consequence of 
the Bose-statistics of the gluon loop.
 
The effective action has a manifest Weyl symmetry,
the six-element subgroup of $SU(3)$ which contains 
the cyclic $Z_3$. 
Furthermore it has the dual symmetry. It is invariant under 
the dual transformation 
$H_p \rightarrow -iE_p$ and $E_p \rightarrow iH_p$.
Notice that this is exactly the same dual symmetry which we have 
in QED and $SU(2)$ QCD \cite{prl01,prd02}. 
We can also express the effective actions (\ref{seaa}) 
and (\ref{seab}) in terms of three
Casimir invariants, $(\hat F_{\mu\nu}^a)^2$,
$(d_{bc}^{~~a} \hat F_{\mu\nu}^b \hat F_{\mu\nu}^c)^2$, and
$(d_{abc} \hat F_{\mu\nu}^a \hat F_{\nu\rho}^b \hat F_{\rho \sigma}^c)^2$,
replacing $H_p$ (and $E_p$) by the Casimir invariants,
which assures the gauge invariance of the effective actions.
But it should be noticed that the imaginary part of the effective 
actions depend only on one Casimir invariant, $(\hat F_{\mu\nu}^a)^2$.

Just as in $SU(2)$ QCD we can obtain the effective potential 
from the effective action. For the constant magnetic background
the effective potential is given by 
\bea
&V_{eff}=\dfrac{1}{2} (\bar H_3^2+\bar H_8^2) +\dfrac{11g^2}{48 \pi^2}
\Big \{\bar H_3^2 \ln \big(\dfrac{g\bar H_3}{\mu^2} -c\big) \nn\\
&+\bar H_+^2 \ln \big(\dfrac{g\bar H_+}{\mu^2} -c\big)
+\bar H_-^2 \ln \big(\dfrac{g\bar H_-}{\mu^2} -c\big), \nn\\
&\bar H_\pm^2=\dfrac{1}{4} \bar H_3^2+\dfrac{3}{4} \bar H_8^2
\pm \dfrac{\sqrt 3}{2} \bar H_3 \bar H_8 \cos \theta, \nn\\
&\bar H_3= \sqrt{(H_{\mu\nu}^3)^2/2},
~~~\bar H_8= \sqrt{(H_{\mu\nu}^8)^2/2}, \nn\\
&\cos \theta = H_{\mu\nu}^3 H_{\mu\nu}^8/2 \bar H_3 \bar H_8.
\label{effpot}
\eea
Notice that the classical potential depends only
on $\bar H_3^2+\bar H_8^2$, but the effective potential 
depends on three variables $\bar H_3$, $\bar H_8$,
and $\cos \theta$. At first thought this might look strange,
but as we have remarked this is because
the effective action depends on {\it three} invariant variables,
which can be chosen to be $\bar H_3$, $\bar H_8$,
and $\cos \theta$. We emphasize that $\cos \theta$ can be arbitrary
because $H_{\mu\nu}^3$ and $H_{\mu\nu}^8$ are completely 
independent, so that they can have different
space polarization. The potential has the unique minimum 
at $\bar H_3=\bar H_8=H_0$ and $\cos \theta =0$.
Notice that when $H_{\mu\nu}^3$ and $H_{\mu\nu}^8$ are 
parallel (or when $\cos \theta=1$) it has two degenerate
minima at $\bar H_3=\tilde H_0,~\bar H_8=0$ and at 
$\bar H_3=\tilde H_0/2,~\bar H_8=\sqrt3 \tilde H_0/2$,
where $\tilde H_0=\root {1/3}\of{2}~H_0$. We plot the effective 
potential for $\cos \theta=1$ in Fig. 1 and for $\cos \theta =0$
in Fig. 2 for comparison.

One can renormalize the potential
by defining a running coupling $\bar g^2(\bar \mu^2)$
\bea
& \dfrac{\pro^2 V_{eff}}{\pro \bar H_i^2} 
\Big |_{\bar H_3=\bar H_8=\bar \mu^2,\theta=\pi/2}
=\dfrac{g^2}{\bar g^2}  \nn \\
&= 1+ \dfrac{11}{16 \pi^2} g^2 (\ln \dfrac{\bar \mu^2}{\mu^2}
-c+\dfrac{5}{4}),~~~~(i=3,8)
\label{renorm}
\eea
from which we retrieve the QCD $\beta$-function 
\bea
& \beta (\bar \mu) = \bar \mu \dfrac{d \bar g}{d \bar \mu}
=-\dfrac{11}{16 \pi^2} \bar g^3.
\label{beta}
\eea
The renormalized potential has the same form 
as in (\ref{effpot}), with the replacement 
$g\rightarrow \bar g,~\mu \rightarrow \bar \mu,~c=5/4$.
It has the unique minimum 
\bea
& V_{min} = -\dfrac{11 \bar \mu^4}{32 \pi^2} \exp \big(-\dfrac{32\pi^2}
{11 \bar g^2} +\dfrac{3}{2} \big), \nn\\
&<\bar H_3>=<\bar H_8>=\dfrac{\bar \mu^2}{\bar g} 
\exp \big(-\dfrac{16\pi^2} {11 \bar g^2} +\dfrac{3}{4} \big).
\eea
It should be noticed that the effective potential
breaks the original $SO(2)$ invariance 
(of $\bar H_3^2+\bar H_8^2$) of the classical Lagrangian.

\begin{figure}
\begin{center}
\psfig{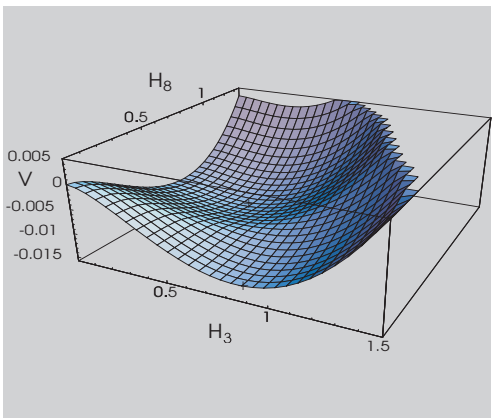}
\end{center}
\caption{\label{Fig. 1} The QCD effective potential with $\cos \theta =1$,
which has two degenerate minima.}
\end{figure}

The QCD effective action has been calculated before 
with different methods \cite{yil,adl}.
Our method has the advantage that it naturally reduces 
the calculation of $SU(N)$ QCD effective action
to that of $SU(2)$ QCD. 

\begin{figure}[t]
\begin{center}
\psfig{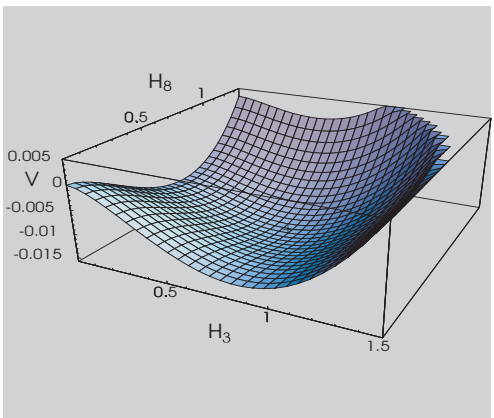}
\end{center}
\caption{\label{Fig. 2} The effective potential with $\cos \theta =0$,
which has a unique minimum at $\bar H_3=\bar H_8=H_0$.}
\end{figure}

There have been a lot of controversies and confusions on 
the imaginary part of the effective 
action \cite{niel,yil,adl,sch,prd02,jhep}.
This is (at least partly) due to the fact that 
the imaginary part depends on the regularization method.
Here we remark that there is a straightforward way to resolve
this controversy. Notice that the imaginary part depends
only on the second order in coupling constant $g$.
This implies that one can calculate the imaginary part 
independently from the perturbative Feynman diagram \cite{sch,jhep}. 
We find that the perturbative calculation
supports the gauge invariant regularization.

Independent of this controversy we emphasize that in both
regularizations the chromoelectric background
generates a negative imaginary part. Only the quarks,
due to the Fermi-statistics, has a positive contribution
to the imaginary part. This should be contrasted with earlier 
results \cite{yil,sch}. 
The detailed discussion of the subject will be published 
elsewhere \cite{cho3}.

{\bf Acknowledgements}
                                                                                
~~~One of the authors (DGP) thanks
N. Kochelev and P. M. Zhang for useful discussions. 
The work is supported in part by the ABRL Program of
Korea Science and Engineering Foundation (R14-2003-012-01002-0) 
and by the BK21 Project of the Ministry of Education.

\end{document}